\documentclass{article}

\usepackage{arxiv}

\usepackage[utf8]{inputenc} 
\usepackage[T1]{fontenc}    
\usepackage{hyperref}       
\usepackage{url}            
\usepackage{booktabs}       
\usepackage{amsfonts}       
\usepackage{nicefrac}       
\usepackage{microtype}      
\usepackage{lipsum}		
\usepackage{graphicx}
\usepackage[square , sort, numbers]{natbib}
\usepackage{doi}
\usepackage[nolist]{acronym}
\usepackage{threeparttable} 
\usepackage{tabularx} 
\usepackage{multirow}
\usepackage{threeparttable}
\usepackage{makecell}
\usepackage{amsmath}
\usepackage{ragged2e}
\usepackage{capt-of}

\title{Improving the Reliability of Anomaly Detection for Encrypted OPC UA Traffic over Private 5G}

\date{September 24, 2026}	

\author{ \href{https://orcid.org/0000-0003-1716-5867}{\includegraphics[scale=0.06]{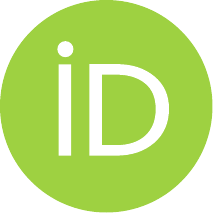}\hspace{1mm}Song Son Ha}\\
	Electrical Measurement Engineering\\
	Helmut-Schmidt-University\\
	Hamburg, Germany\\
	\texttt{song.ha@hsu-hh.de} \\
	\And
	\href{https://orcid.org/0009-0003-4437-010X}{\includegraphics[scale=0.06]{orcid.pdf}\hspace{1mm}Florian Foerster}\\
	Institute for Innovative Safety and Security\\
	Technical University of Applied Sciences Augsburg\\
	Augsburg, Germany\\
	\texttt{florian.foerster@tha.de} \\
	\And
	\href{https://orcid.org/0000-0002-5390-3946}{\includegraphics[scale=0.06]{orcid.pdf}\hspace{1mm}Henry Beuster}\\
	Electrical Measurement Engineering\\
	Helmut-Schmidt-University\\
	Hamburg, Germany\\
	\texttt{henry.beuster@hsu-hh.de} \\
	\And
	{Tim Kittel}\\
	ipoque GmbH\\
	A Rohde \& Schwarz company\\
	Leipzig, Germany\\
	\texttt{tim.kittel@rohde-schwarz.com} \\
	\And
	\href{https://orcid.org/0000-0003-2310-5895}{\includegraphics[scale=0.06]{orcid.pdf}\hspace{1mm}Dominik Merli}\\
	Institute for Innovative Safety and Security\\
	Technical University of Applied Sciences Augsburg\\
	Augsburg, Germany\\
	\texttt{dominik.merli@tha.de} \\
	\And
	{Gerd Scholl}\\
	Electrical Measurement Engineering\\
	Helmut-Schmidt-University\\
	Hamburg, Germany\\
	\texttt{gerd.scholl@hsu-hh.de} \\
}

\renewcommand{\shorttitle}{Reliable Anomaly Detection for Encrypted OPC UA over Private 5G}

\hypersetup{
	hidelinks,
	pdftitle={Improving the Reliability of Anomaly Detection for Encrypted OPC UA Traffic over Private 5G},
	pdfsubject={Improving the Reliability of Anomaly Detection for Encrypted OPC UA Traffic over Private 5G},
	pdfauthor={Song Son Ha; Florian Foerster; Henry Beuster; Tim Kittel; Dominik Merli; Gerd Scholl},
	pdfkeywords={OPC~UA, Encrypted Traffic Analysis, Anomaly Detection, Private~5G, Control-Plane Context}
}

\begin{document}
	\maketitle

\begingroup
\renewcommand{\thefootnote}{}
\footnotetext{%
	This is the authors' accepted version of a paper to be presented at the
	19th IEEE International Conference on Cyber, Physical and Social Computing
	(CPSCom 2026), Montbéliard, France, November 30--December 4, 2026.	
	\copyright~2026 IEEE. Personal use of this material is permitted.
	Permission from IEEE must be obtained for all other uses, including
	reprinting/republishing this material for advertising or promotional purposes,
	creating new collective works, for resale or redistribution to servers or lists,
	or reuse of any copyrighted component of this work in other works.}
\addtocounter{footnote}{-1}
\endgroup
	
\begin{abstract}
	
Open Platform Communications Unified Architecture (OPC~UA) is increasingly deployed over private~5G networks in industrial environments, where end-to-end encryption prevents payload inspection by network-based intrusion detection systems (IDSs). Although payload-agnostic statistical features extracted from encrypted traffic enable traffic-based anomaly detection, benign connectivity variations may alter observable user-plane (UP) behavior and increase the false-positive rate (FPR). This paper investigates this reliability problem and proposes a control-plane (CP)-aware decision adaptation for four frozen IDS models. CP indicators at the user equipment (UE) level are used to construct a temporal CP context in which a CP-specific threshold selected on adaptation validation data is applied, while the original threshold remains active outside the context. The traffic features, attack scores, preprocessing procedure, and trained model parameters remain unchanged.

Evaluation on a real industrial private~5G testbed shows that benign connectivity variations increase the FPR for all four evaluated models and that false positives are concentrated within periods temporally associated with CP activity. The proposed CP-aware decision adaptation reduces both global FPR and FPR within the CP context while introducing a configurable trade-off between FPR reduction and retained recall over the complete attack campaign. These findings demonstrate that CP context can improve the operational reliability of encrypted-traffic intrusion detection without retraining the underlying models.
	
\end{abstract}


\section{Introduction}
\label{sec:introduction}

Open Platform Communications Unified Architecture (OPC~UA) is widely used in industrial automation because of its interoperability, extensibility, and integrated security mechanisms~\cite{honda2022monitoring}. Private~5G networks provide dedicated wireless connectivity and quality-of-service control for heterogeneous industrial devices and applications. Their combination supports flexible wireless industrial communication, but end-to-end OPC~UA encryption limits network-side security monitoring. When using the \texttt{SignAndEncrypt} security mode, application payloads and service semantics are not accessible at the monitoring point. Intrusion detection must therefore rely on remaining payload-agnostic traffic characteristics such as packet sizes, timing, directionality, flow dynamics, and observable protocol framing.

Traffic-based intrusion detection system (IDS) models are commonly calibrated and evaluated under relatively stable connectivity conditions. Benign user equipment (UE) reconnections, Protocol Data Unit (PDU) session changes, and temporary communication interruptions may alter observable user-plane (UP) traffic without indicating malicious activity. The resulting changes in traffic intensity, timing, and session behavior can resemble application anomalies and increase the false-positive rate (FPR).

Previous work~\cite{ha2026impact} experimentally demonstrated that benign private~5G connectivity variations increase the FPR of IDS models for encrypted OPC~UA traffic. Elevated attack scores also occurred frequently during or shortly after UE-level control-plane (CP) activity. That study used CP indicators only to interpret IDS outputs and did not introduce a mechanism for adapting IDS decisions. It also did not evaluate the effect of such adaptation when attacks and benign connectivity variations occur within the same capture.

This paper extends the preceding study from temporal interpretation to decision adaptation. Lightweight UE-level CP indicators are used to construct a temporal CP context associated with connectivity activity. The adaptation operates only at the decision stage of the frozen IDS models. CP-specific thresholds are selected using benign variation adaptation validation data to reduce FPR and mixed variation--attack adaptation validation data to constrain the reduction in recall over the complete attack campaign.

The main contributions of this work are as follows:
\begin{itemize}
	\item A systematic quantification of false-positive concentration within and outside the temporal CP context across four frozen IDS models under controlled benign connectivity variations.
	
	\item A CP-aware decision adaptation that applies a CP-specific threshold only within the temporal CP context while preserving the underlying traffic-based IDS.
	
	\item A validation procedure that selects CP-specific operating points under explicit constraints on recall over the attack campaign, followed by a separate test evaluation using benign variation and mixed variation--attack captures.
\end{itemize}

The remainder of this paper is organized as follows. Section~\ref{sec:related_work} reviews related work. Section~\ref{sec:methodology} presents the proposed methodology. Section~\ref{sec:implementation} describes the experimental implementation and dataset construction. Section~\ref{sec:results} reports the evaluation results. Section~\ref{sec:conclusion} concludes the paper and outlines directions for future work.

\section{Related Work}
\label{sec:related_work}

Research related to this work covers encrypted traffic analysis, intrusion detection for encrypted OPC~UA communication, analysis of 5G control-plane behavior, and the use of operational context in anomaly detection.

Encrypted traffic analysis commonly relies on statistical characteristics observable without application-payload access, including packet sizes, timing, directionality, and flow statistics~\cite{shen2023encryptedsurvey,jisensors2024slr}.
Such representations can change across deployment conditions: Malekghaini et al.~\cite{malekghaini2023drift} report degradation of encrypted-traffic classifiers under real-world data drift, while Qing et al.~\cite{qing2025dynamicencrypted} study robust classification under dynamic network conditions.
Although classification-focused, these studies show that traffic-derived model behavior need not remain stable after deployment.

For OPC~UA, Honda et al.~\cite{honda2021encryptedopcua,honda2022monitoring} address the loss of direct visibility caused by encryption through an intermediary monitoring architecture that decrypts the communication.
Neu et al.~\cite{neu2019opcua} investigate attacks by untrusted OPC~UA clients, including encrypted denial-of-service traffic, and corresponding detection approaches.
In the broader industrial domain, Wu et al.~\cite{wu2025fread} propose anomaly detection for encrypted ICS traffic.
Statistical features extracted from encrypted OPC~UA traffic have also been shown to support the detection of protocol-compliant attacks over industrial private~5G~\cite{ha2026residual}.
These studies establish the feasibility of security monitoring despite encrypted industrial traffic, but do not examine how benign private~5G connectivity procedures affect the operating point of a frozen detector.

Several studies analyze anomaly detection in the 5G control plane using Non-Access Stratum (NAS), NG Application Protocol (NGAP), and core-network signaling sequences~\cite{fritznoms2025granomaly,wangicpads2023gsad,tianictc2023adseq}.
Their objective is to identify abnormal signaling behavior, mobility anomalies, or irregular session procedures within the 5G infrastructure.
Here, CP indicators are neither classified nor included in the traffic-based IDS input; they provide temporal context for encrypted user-plane traffic.
MobiFlow~\cite{wen2022mobiflow} combines UE state, control-plane message sequences, user-plane traffic, and connection-related statistics for security analysis.
Abdelrahman and Gelenbe use network KPIs and data-plane traffic measurements to detect control-plane anomalies in mobile networks~\cite{abdelrahman2016dataplane}.
Neither approach uses benign UE-level CP activity to gate the decision threshold of an otherwise frozen user-plane IDS.

Context-aware anomaly detection uses auxiliary information to distinguish abnormal observations from legitimate operating changes.
Existing approaches incorporate context into feature representations, model structures, or online adaptation~\cite{stodt2026context,toor2025uocad2,yasaei2020iotcad}.
Threshold selection is an important component of network anomaly detection, with supervised and unsupervised strategies providing different approaches to deriving decision thresholds~\cite{komadina2024thresholds}.
At the decision stage, Ali et al.~\cite{ali2013adaptivethreshold} adapt anomaly thresholds under time-varying network and host inputs, Gamallo-Fern\'andez et al.~\cite{gamallofernandez2026contextthresholds} define separate thresholds for different execution conditions, and \.{Z}elasko~\cite{zelasko2026costaware} formulates validation-based operating points that may be indexed by operational context in a network IDS.
These works establish adaptive and context-dependent thresholding as general mechanisms, but do not use transient UE-level 5G CP activity to activate a secondary threshold for an otherwise unchanged encrypted application-traffic detector.

The preceding study~\cite{ha2026impact} further showed that benign 5G connectivity variations increase the FPR of encrypted OPC~UA traffic-based IDS models and that elevated attack scores are temporally associated with CP activity, but used this information only for interpretation. The present study extends this analysis by using UE-level CP activity as temporal context for decision adaptation, selecting CP-specific operating points under an explicit campaign recall constraint, and evaluating them on mixed variation--attack captures.

\section{Proposed Methodology}
\label{sec:methodology}

This section describes the proposed CP-aware decision adaptation for encrypted OPC~UA traffic over private~5G.

\subsection{Overview of the Proposed CP-Aware Decision Adaptation}
\label{subsec:work_flow}

The proposed adaptation combines traffic-based anomaly detection with CP context at the decision stage. Encrypted UP traffic is summarized over fixed time windows (TWs) and processed by a frozen IDS model to obtain an attack score for each TW. In parallel, UE-level CP indicators are aligned with the same TWs to construct a temporal CP context representing periods of connectivity activity. A CP-specific threshold selected during adaptation validation is applied within the CP context, while the original threshold remains active outside it. The CP information is therefore used only at the decision stage and does not affect the attack scores produced by the frozen IDS model.

\subsection{Threat Model}
\label{subsec:threat_model}

This work considers a secured private~5G environment with encrypted and authenticated OPC~UA communication~\cite{vom2022opc}. All sessions use the \texttt{SignAndEncrypt} security mode, preventing the network-based IDS from inspecting plaintext application payloads and service semantics. An adversary may control or compromise an authorized OPC~UA endpoint and generate syntactically valid malicious behavior within encrypted sessions without breaking the cryptographic mechanisms. The resulting attacks are therefore reflected in observable traffic dynamics rather than malformed packets or explicit protocol violations. The considered 5G connectivity variations are benign and are not initiated by the adversary. The UE-level CP telemetry used for the adaptation is assumed to be trusted and correctly time-aligned with the UP observations. Adversarial manipulation of CP telemetry or intentional triggering of CP activity is outside the present threat model.

\subsection{Traffic Representation and Frozen IDS Models}
\label{subsec:traffic_features}
\label{subsec:model_calibration}

Each capture is divided into consecutive, non-overlapping 5\,s TWs. For every TW, bidirectional OPC~UA communication is represented using traffic volume and direction, temporal and flow dynamics, and observable OPC~UA framing and lifecycle activity~\cite{ha2026residual}. Flow-level measurements are aggregated into one feature vector per TW.

The evaluation considers Logistic Regression (LogReg), Random Forest (RF), Support Vector Machine with a radial basis function kernel (SVM), and Extreme Gradient Boosting (XGBoost). Model development and baseline evaluation are described in~\cite{ha2026residual}. The traffic representation, preprocessing procedure, model parameters, score functions, and original decision thresholds $\tau_m$ remain fixed throughout the evaluation.	The benign variation and mixed variation--attack datasets are not used to retrain the models or recalibrate their original thresholds. CP indicators are processed separately and affect only the decision adaptation.

\begin{table}[t]
	\centering
	\caption{UE-level CP indicators extracted from NAS and NGAP traces}
	\label{tab:cp_context}
	\renewcommand{\arraystretch}{1.08}
	\begin{threeparttable}
		\footnotesize
		\begin{tabular}{
				@{}
				>{\raggedright\arraybackslash}p{0.22\columnwidth}
				@{\hspace{5pt}}
				>{\raggedright\arraybackslash}p{0.75\columnwidth}
				@{}
			}
			\toprule
			\textbf{CP Indicator} & \textbf{Description} \\
			\midrule
			
			Registration Event &
			Binary flag indicating the occurrence of a 5G registration or deregistration procedure. \\
			
			\addlinespace[3pt]
			
			CM State Transition &
			Binary flag indicating signaling associated with transitions between CM-IDLE and CM-CONNECTED, derived from NAS and NGAP. \\
			
			\addlinespace[3pt]
			
			PDU Session State Change &
			Binary flag indicating PDU session establishment or release activity. \\
			
			\addlinespace[3pt]
			
			PDU Resource Reconfiguration &
			Binary flag indicating an NGAP PDU session resource setup, modify, or release procedure. \\
			
			\addlinespace[3pt]
			
			CP Signaling Count &
			Aggregate number of CP signaling frames containing NAS-5GS or NGAP information within the TW. \\
			
			\bottomrule
		\end{tabular}
	\end{threeparttable}
\end{table}

\subsection{CP Context Indicators}
\label{subsec:cp_context}

Benign connectivity procedures may temporarily alter UP communication and increase the FPR of traffic-based IDS models. The proposed method uses UE-level CP signaling to identify the time intervals associated with these procedures.	The CP indicators are derived from NAS and NGAP messages collected by the core network operations and monitoring framework. Only signaling associated with the UE subjected to the connectivity variation is processed. In captures containing both a connectivity variation and an attack, the CP indicators are extracted from the UE subjected to the variation, which is different from the attacked UE.

Table~\ref{tab:cp_context} summarizes the selected indicators. The procedure-specific indicators represent registration activity, Connection Management (CM) state transitions, Protocol Data Unit (PDU) session activity, and access-side PDU resource procedures. The aggregate CP signaling count provides a procedure-agnostic indication of UE-level CP activity. It therefore preserves CP activity that is not mapped to one of the selected procedure-specific indicators and also provides a consistency check for their extraction.

For each TW $t$, the CP indicators form a context vector $\mathbf{c}_t$. A TW is marked as CP-active when at least one component of $\mathbf{c}_t$ is nonzero. CP-active TWs that are temporally close are merged into CP activity clusters using a maximum internal gap. The temporal CP context is then obtained by extending the resulting clusters with configurable pre- and post-padding. The gap and padding parameters are selected exclusively on the benign variation adaptation validation captures and are fixed before test evaluation.

\subsection{CP-Aware Decision Adaptation}
\label{subsec:context_adaptation}

The proposed adaptation changes only the decision threshold. For IDS model $m$ and TW $t$, let $s_{m,t}\in\mathbb{R}$ denote the attack score and let $\tau_m$ denote the original threshold of the frozen model. Thresholds are handled separately because the ranges of the attack scores differ across models. The original decision is

\begin{equation}
	d_{m,t} =
	\begin{cases}
		1, & s_{m,t} \geq \tau_m,\\
		0, & s_{m,t} < \tau_m,
	\end{cases}
	\label{eq:baseline_decision}
\end{equation}

where $d_{m,t}=1$ denotes an attack prediction. The CP indicators are not included in the model input and do not affect $s_{m,t}$.

\subsubsection{Decision Rule with CP Context}

Let $q_t\in\{0,1\}$ indicate whether TW $t$ belongs to the temporal CP context defined in Section~\ref{subsec:cp_context}. A CP-specific threshold $\tau_{m,\mathrm{CP}}$ is applied when $q_t=1$, while the original threshold $\tau_m$ remains active when $q_t=0$. The threshold used at TW $t$ is

\begin{equation}
	\tau_{m,t}^{\mathrm{CP}}
	=
	(1-q_t)\tau_m
	+
	q_t\tau_{m,\mathrm{CP}}.
	\label{eq:context_threshold}
\end{equation}

The adapted decision is

\begin{equation}
	\tilde{d}_{m,t} =
	\begin{cases}
		1, & s_{m,t} \geq \tau_{m,t}^{\mathrm{CP}},\\
		0, & s_{m,t} < \tau_{m,t}^{\mathrm{CP}}.
	\end{cases}
	\label{eq:cp_adapted_decision}
\end{equation}

Only thresholds satisfying $\tau_{m,\mathrm{CP}}\geq\tau_m$ are evaluated. The adaptation can suppress attack predictions within the CP context but cannot introduce new ones. Decisions outside the context remain unchanged.

The CP context is not interpreted as evidence that a TW is benign. It identifies intervals in which benign connectivity procedures may alter UP traffic and increase false positives. Genuine attacks may also occur within these intervals. The CP-specific threshold is therefore selected under a constraint on recall over the complete attack campaign.

\subsubsection{Construction of Tested Thresholds}

For model $m$, let $\lambda$ denote one threshold value to be tested within the CP context. During the evaluation of $\lambda$, the decision rule in~\eqref{eq:context_threshold} is applied by setting $\tau_{m,\mathrm{CP}}=\lambda$. The possible values of $\lambda$ are derived from the attack scores $s_{m,t}$ of TWs within the CP context whose scores satisfy $s_{m,t}\geq\tau_m$ under the original decision rule. On the benign variation adaptation validation set, every TW is benign. Therefore, each selected score $s_{m,t}$ directly corresponds to a false-positive decision made by the original threshold.

On the mixed variation--attack adaptation validation set, the selected scores $s_{m,t}$ belong to evaluated TWs within the attack campaign that satisfy $s_{m,t}\geq\tau_m$ and therefore correspond to correct attack predictions. Increasing the CP-specific threshold may suppress the false positives represented by the first group, but may also remove correct attack predictions represented by the second group.

Let $\mathcal{S}_m$ denote the set of attack scores $s_{m,t}$ obtained from these two groups of TWs. Repeated score values are considered only once. For each $s_{m,t}\in\mathcal{S}_m$, the procedure tests the smallest representable threshold $s_{m,t}^{+}>s_{m,t}$. When $\lambda=s_{m,t}^{+}$ is applied within the CP context, TWs with score $s_{m,t}$ are classified as benign because $s_{m,t}<\lambda$.

The set of tested thresholds is

\begin{equation}
	\mathcal{C}_m
	=
	\{\tau_m\}
	\cup
	\left\{
	s_{m,t}^{+}
	\;\middle|\;
	s_{m,t}\in\mathcal{S}_m
	\right\}.
	\label{eq:candidate_threshold_set}
\end{equation}

Including $\tau_m$ allows the original decision rule to remain a possible operating point. Every threshold $\lambda\in\mathcal{C}_m$ satisfies $\lambda\geq\tau_m$, so the adaptation can suppress attack predictions within the CP context but cannot introduce additional ones.

\subsubsection{Selection of Operating Points}

For each threshold $\lambda\in\mathcal{C}_m$, the global FPR is measured on the benign variation adaptation validation set, while the corresponding reduction in campaign recall is measured on the mixed variation--attack adaptation validation set. Each $\lambda$ is evaluated as a possible value of $\tau_{m,\mathrm{CP}}$. For a given tolerance $\delta_{\mathrm{val}}$, the selected value is denoted by $\tau_{m,\mathrm{CP}}^{(\delta_{\mathrm{val}})}$.

Each mixed variation--attack capture contains repeated attack bursts separated by inter-burst intervals. The complete attack campaign extends from the first to the final attack-burst TW, including the intervening inter-burst TWs.

Let $R_{m,\mathrm{camp}}^{(\tau_m)}$ denote recall over these TWs when the original threshold is used both within and outside the CP context. Let $R_{m,\mathrm{camp}}^{(\lambda)}$ denote the corresponding recall when $\lambda$ is applied within the CP context and $\tau_m$ remains active outside it. The campaign recall reduction on the adaptation validation set, expressed in percentage points (pp), is

\begin{equation}
	\Delta R_{m,\mathrm{camp}}^{(\lambda)}
	=
	100
	\left(
	R_{m,\mathrm{camp}}^{(\tau_m)}
	-
	R_{m,\mathrm{camp}}^{(\lambda)}
	\right).
	\label{eq:campaign_recall_drop}
\end{equation}

The evaluated tolerances are

\begin{equation}
	\mathcal{D}
	=
	\{0,\;0.5,\;1,\;2,\;5\}\ \mathrm{pp},
	\label{eq:validation_tolerances}
\end{equation}

For each tolerance $\delta_{\mathrm{val}}\in\mathcal{D}$, the selected CP-specific threshold is

\begin{equation}
	\begin{aligned}
		\tau_{m,\mathrm{CP}}^{(\delta_{\mathrm{val}})}
		&=
		\underset{\lambda\in\mathcal{C}_m}{\arg\min}
		\;\mathrm{FPR}_{m,\mathrm{var}}^{(\lambda)}
		\\[-1mm]
		&\quad \text{subject to}\quad
		\Delta R_{m,\mathrm{camp}}^{(\lambda)}
		\leq \delta_{\mathrm{val}}.
	\end{aligned}
	\label{eq:operating_point_selection}
\end{equation}

Here, $\mathrm{FPR}_{m,\mathrm{var}}^{(\lambda)}$ denotes the global FPR measured on the benign variation adaptation validation set when $\lambda$ is applied within the CP context. From $\mathcal{C}_m$, the procedure selects the threshold that produces the lowest validation FPR while satisfying the permitted campaign recall reduction. As an additional validation safeguard, candidate thresholds that reduce attack-burst recall by more than the same tolerance $\delta_{\mathrm{val}}$ are excluded.

A tolerance of $0$~pp requires campaign recall to remain unchanged on the mixed variation--attack adaptation validation set, while the attack-burst safeguard prevents a recall reduction during attack bursts. This condition does not require $\tau_{m,\mathrm{CP}}^{(\delta_{\mathrm{val}})}=\tau_m$, because a higher threshold may suppress false positives without violating the validation constraints. Larger tolerances permit greater recall reductions and may therefore provide a larger FPR reduction.

If several thresholds produce the same minimum FPR, preference is given to the threshold with the smaller maximum reduction across campaign recall and attack-burst recall. Remaining ties are resolved using the smaller inter-burst recall reduction and then the smaller threshold increase.

\begin{figure}[b!]
	\centering
	\includegraphics[width=0.6\linewidth]{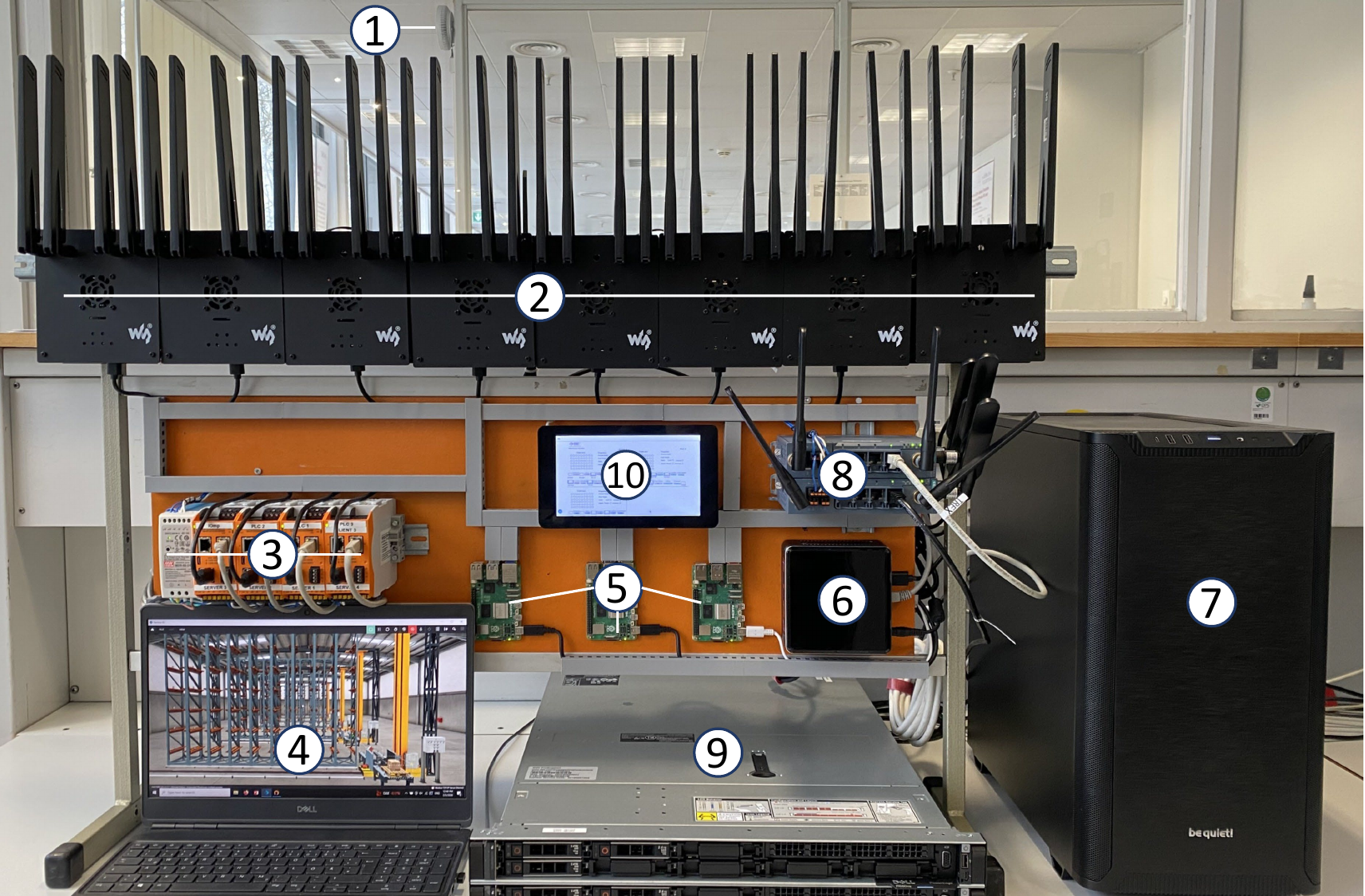}
	\caption{Experimental testbed:
		1) 5G radio units;
		2) SIMCom SIM8202G-M2 modules;
		3) Revolution~Pi modules;
		4) Factory~I/O system;
		5) Raspberry~Pi~5 modules;
		6) Intel~NUC;
		7) attack workstation;
		8) Siemens Scalance~M800 router;
		9) IDS backend; and
		10) testbed orchestrator.
		The 5G core network, central and distributed units, telemetry services, and remote engineering systems are not shown.}
	\label{fig:testbed}
\end{figure}

\subsubsection{Test Evaluation}

The complete set of operating points selected for the five validation tolerances is fixed before test evaluation. Global FPR and FPR within the CP context are measured on the benign variation test captures. Campaign recall is measured separately on the mixed variation--attack test captures. FPR reduction is reported relative to the corresponding FPR under the original decision rule, while campaign recall reduction is reported in pp. The FPR outside the CP context remains unchanged because the original threshold $\tau_m$ is retained in these TWs. No test capture is used to select the CP-context parameters, construct $\mathcal{C}_m$, or select an operating point. The validation constraints are applied only during adaptation validation. Campaign recall reduction on the mixed variation--attack test captures may exceed the corresponding tolerance and does not trigger threshold reselection.

\section{Implementation}
\label{sec:implementation}

\subsection{Experimental Environment}
\label{subsec:experimental_environment}

The experiments are conducted on the industrial private~5G testbed described in~\cite{11205743}. The OPC~UA application operates under the encrypted configuration described in~\cite{ha2026residual}. Only the components relevant to the present data collection are summarized here. The platform provides a standalone network compliant with 3GPP Release~16, mirrored UP traffic, and UE-level CP traces. A warehouse automation process is emulated using Factory~I/O~\cite{factoryio}. Revolution~Pi controllers execute the programmable logic controller functions and host OPC~UA endpoints. Human-machine interface, supervisory control and data acquisition, engineering, diagnostic, and remote controller systems generate the industrial communication workloads. All OPC~UA sessions use the \texttt{SignAndEncrypt} security mode with the \texttt{Aes128-Sha256-RsaOaep} security policy. Mirrored UP traffic is transferred to the IDS backend, where the R\&S\textsuperscript{\textregistered}PACE~2 library extracts observable traffic and OPC~UA framing information without decrypting service payloads. The resulting observations are aggregated into TWs. CP traces are collected separately for the relevant UE and aligned with the UP observations during dataset construction. Fig.~\ref{fig:testbed} shows the physical components of the testbed.

\subsection{Attack Scenarios}
\label{subsec:attacks}

The attack implementations extend the OPC~UA exploitation framework provided by Claroty~\cite{claroty2025} to operate through authorized OPC~UA clients under the \texttt{SignAndEncrypt} security mode. The evaluated families are Browse Address Space, Publish Request Flooding, Persistent Secure Channel Exhaustion, and Translate Browse Path, targeting address-space traversal, subscription processing, SecureChannel and session resources, and browse-path resolution, respectively. Each family is executed at four configuration levels using repeated attack bursts separated by inter-burst intervals. The parameter settings and configuration progression are described in~\cite{ha2026residual}.

\subsection{Controlled Connectivity Variation Scenarios}
\label{subsec:env_variation}

The evaluation considers four controlled benign connectivity variations. Each variation is triggered on a selected UE and invokes a specific private~5G connectivity procedure without introducing malicious application behavior. The scenarios are designed to produce representative changes in CP signaling and UP communication that may affect the decisions of the frozen traffic-based IDS models. Table~\ref{tab:env_variations} summarizes the implemented procedures, their expected effects, and representative industrial contexts.

\begin{table}[b!]
	\centering
	\caption{Overview of controlled benign connectivity variation scenarios}
	\label{tab:env_variations}
	\renewcommand{\arraystretch}{1.3}
	\begin{tabularx}{\textwidth}{@{} >{\raggedright\arraybackslash}p{1.7cm} X p{6.1cm} p{3.9cm} @{}}
		\toprule
		\textbf{Scenario} & \textbf{Description} & \textbf{Expected Effects} & \textbf{Example Industrial Context} \\
		\midrule
		
		Hard UE Reconnection &
		Complete modem restart followed by reinitialization of the UE communication stack. &
		UE registration, security procedures, and PDU session re-establishment, accompanied by a pronounced CP signaling burst and temporary UP interruption. &
		Device reboot, modem recovery, or power-cycle event. \\
		
		Soft UE Reconnection &
		UE-initiated NAS deregistration followed by re-registration without restarting the modem. &
		Deregistration and registration signaling followed by PDU session re-establishment and temporary UP interruption. &
		Software-triggered reconnection, communication stack restart, or remote maintenance. \\
		
		PDU Session Reset &
		Release and subsequent re-establishment of the PDU session while UE registration is retained. &
		PDU session management and NGAP resource signaling, with temporary loss and restoration of UP connectivity. &
		Session recovery after maintenance or a transient data-path fault. \\
		
		PDU Session Interruption &
		Temporary interruption of UP connectivity without a complete UE reconnection or explicit PDU session reset. &
		Temporary absence of UP communication and possible CM or PDU resource signaling during interruption and recovery. &
		Temporary radio degradation, link outage, or short communication-path disruption. \\
		
		\bottomrule
	\end{tabularx}
\end{table}

\subsection{Dataset Construction and Partitioning}
\label{subsec:dataset_construction}
The dataset comprises 522 captures with a nominal duration of 10 min each, corresponding to a nominal total of 87\,h of traffic. Actual capture durations vary slightly depending on the completion of the experimental procedure. All captures are partitioned before extraction of the non-overlapping 5\,s TWs, and no capture appears in more than one subset. Table~\ref{tab:dataset_summary} summarizes the capture-level partitioning.

The model development data comprise 90 stable benign captures and 160 malicious-only captures. The malicious captures cover four attack families, four configuration levels, and ten repetitions for each family and level combination. Five repetitions are assigned to model training, two to model validation, and three to testing.

The benign variation data extend the collection reported in the preceding study~\cite{ha2026impact} from 40 to 80 captures by adding ten new captures for each of the four connectivity scenarios. Each capture contains two controlled executions of the corresponding connectivity variation. Five captures per scenario are assigned to adaptation validation, and the remaining 15 are reserved for testing.

The mixed variation--attack data comprise all combinations of four connectivity scenarios, four attack families, four attack levels, and three repetitions. One repetition from each combination is assigned to adaptation validation, and two are reserved for testing. The connectivity variation and the attack affect different UEs in every mixed capture. CP indicators are extracted from the signaling trace associated with the UE subjected to the connectivity variation.

\begin{table}[t]
	\centering
	\caption{Capture-level dataset partitioning}
	\label{tab:dataset_summary}
	\renewcommand{\arraystretch}{1.12}
	\setlength{\tabcolsep}{3.5pt}
	\small
	
	\begin{tabular}{@{} lcccccc @{}}
		\toprule
		\multirow{2}{*}{\textbf{Traffic category}} &
		\multicolumn{2}{c}{\textbf{Model}} &
		\textbf{Adapt.} &
		\multirow{2}{*}{\textbf{Test}} &
		\multirow{2}{*}{\textbf{Total}} &
		\multirow{2}{*}{\textbf{Nom. dur.}} \\
		\cmidrule(lr){2-3}
		&
		\textbf{Train} &
		\textbf{Val.} &
		\textbf{Val.} &
		&
		&
		\\
		\midrule
		
		Stable benign      & 63 & 9  & 0  & 18  & 90  & 15.0\,h \\
		Malicious-only     & 80 & 32 & 0  & 48  & 160 & 26.7\,h \\
		Benign variation   & 0  & 0  & 20 & 60  & 80  & 13.3\,h \\
		Mixed variation--attack  & 0  & 0  & 64 & 128 & 192 & 32.0\,h \\
		\midrule
		
		\textbf{Overall total} &
		\textbf{143} &
		\textbf{41} &
		\textbf{84} &
		\textbf{254} &
		\textbf{522} &
		\textbf{87.0\,h} \\
		\bottomrule
	\end{tabular}
	
	\vspace{1mm}
	\parbox{\linewidth}{\footnotesize
		Values without units denote numbers of captures.
		\textit{Adapt.}, \textit{Val.}, and \textit{Nom. dur.} denote adaptation,
		validation, and nominal duration, respectively.}
\end{table}

\subsection{CP Context Parameter Selection}
\label{subsec:cp_context_parameters}

The CP context parameters were determined using only the benign variation adaptation validation captures. A sensitivity analysis showed that an internal gap of five TWs was the smallest tested value for which the number of CP activity clusters stabilized. The cluster count remained unchanged for all larger tested gaps. The padding parameters were selected by jointly considering the mean fraction of false positives under the original decision rule covered by the CP context across the four IDS models and the fraction of TWs included in the context. Increasing the post-padding substantially increased false-positive coverage up to seven TWs, while the increase from seven to eight TWs provided only a limited additional gain. A post-padding of seven TWs was therefore selected. With this value fixed, additional pre-padding provided comparatively limited gains in false-positive coverage while further increasing the fraction of TWs included in the context. Pre-padding was therefore set to zero. These parameters were subsequently fixed for all test evaluations.

\section{Evaluation Results}
\label{sec:results}

The four traffic-based IDS models and their original thresholds remain fixed throughout the evaluation. For reference, the frozen models yield FPRs between 4.60\% and 4.99\% on the stable benign test set from the preceding study~\cite{ha2026residual}. The following analysis examines the increase in FPR under benign connectivity variations, differences in FPR within and outside the CP context, and the trade-off between FPR reduction and retained campaign recall under the proposed CP-aware decision adaptation.

\subsection{Impact of Benign Connectivity Variations}

\subsubsection{FPR under Benign Variations}

\begin{figure}[t]
	\centering
	\includegraphics[width=0.7\columnwidth]{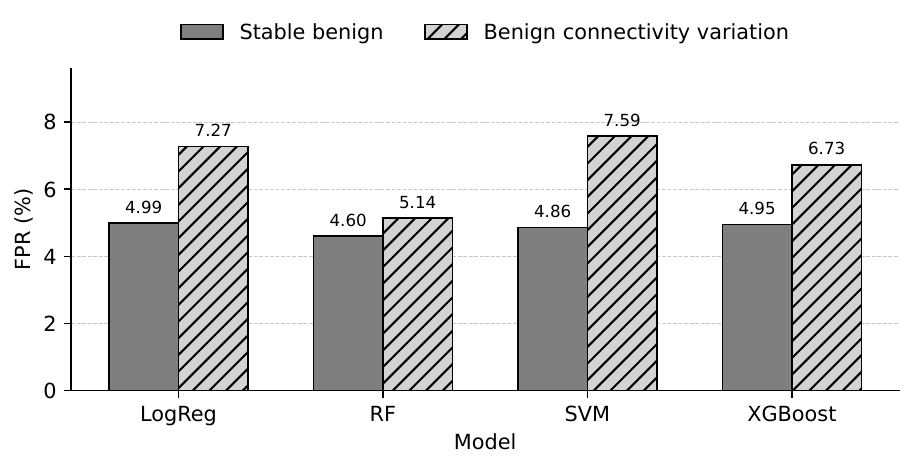}
	\caption{FPR under stable benign conditions and benign connectivity variations.}
	\label{fig:fpr_shift_benign_variations}
\end{figure}

Figure~\ref{fig:fpr_shift_benign_variations} compares the FPR of each frozen IDS model on the stable benign test set and the benign variation test set. Benign connectivity variations increase the FPR for all four models. The FPR rises from 4.99\% to 7.27\% for LogReg, from 4.60\% to 5.14\% for RF, from 4.86\% to 7.59\% for SVM, and from 4.95\% to 6.73\% for XGBoost. The corresponding increases range from 0.54~pp for RF to 2.73~pp for SVM. These results show that legitimate connectivity procedures can shift encrypted UP traffic sufficiently to produce additional false positives.

\begin{table}[t]
	\centering
	\caption{FPR within and outside the CP context on the benign variation test set.}
	\label{tab:fpr_cp_context}
	\setlength{\tabcolsep}{5pt}
	\small
	\renewcommand{\arraystretch}{1.05}
	\begin{threeparttable}
		\begin{tabular}{lccc}
			\toprule
			\multirow{2}{*}{\textbf{Model}}
			& \textbf{FPR outside}
			& \textbf{FPR inside}
			& \textbf{Relative FPR} \\
			& \textbf{CP context [\%]}
			& \textbf{CP context [\%]}
			& \textbf{increase [\%]} \\
			\midrule
			LogReg   & 5.96 & 10.79 & 81.0 \\
			RF       & 4.47 & 6.94 & 55.0 \\
			SVM      & 5.42 & 13.42 & 147.5 \\
			XGBoost  & 5.40 & 10.29 & 90.4 \\
			\bottomrule
		\end{tabular}
		\begin{tablenotes}[flushleft]
			\footnotesize
			\item Relative FPR increases are calculated from unrounded values using the FPR outside the CP context as the reference.
		\end{tablenotes}
	\end{threeparttable}
\end{table}

\begin{table}[t]
	\centering
	\caption{FPR and campaign recall trade-off on the test sets for CP-aware operating points selected on adaptation validation data}
	\label{tab:cp_adaptation_tradeoff}
	\setlength{\tabcolsep}{3pt}
	\begin{threeparttable}
		\footnotesize
		\begin{tabular}{lccc@{\hspace{15pt}}c@{\hspace{15pt}}ccc@{\hspace{10pt}}cc@{\hspace{-2pt}}c}
			\toprule
			\multirow{2}{*}{\textbf{Model}} &
			\multirow{2}{*}{\makecell{\(\boldsymbol{\delta_{\mathrm{val}}}\)\\\textbf{(pp)}}} &
			\multirow{2}{*}{\makecell{\(\boldsymbol{\Delta R_{m,\mathrm{camp}}^{\mathrm{test}}}\)\\\textbf{(pp)}}} &
			\multicolumn{2}{c}{\textbf{Threshold}} &
			\multicolumn{3}{c}{\textbf{Global FPR (\%)}} &
			\multicolumn{3}{c}{\hspace{-10pt}\textbf{FPR within the CP context (\%)}} \\
			\cmidrule(l{3pt}r{10pt}){4-5}
			\cmidrule(l{1pt}r{12pt}){6-8}
			\cmidrule(l{1pt}r{12pt}){9-11}
			& & &
			\(\boldsymbol{\tau_m}\) &
			\(\boldsymbol{\tau_{m,\mathrm{CP}}^{(\delta_{\mathrm{val}})}}\) &
			\textbf{Original} &
			\textbf{Adapted} &
			\makecell{\textbf{RR} \textbf{(\%)}} &
			\textbf{Original} &
			\textbf{Adapted} &
			\makecell{\textbf{RR} \textbf{(\%)}} \\
			\midrule
			
			\multirow{5}{*}{LogReg}
			& 0   & 0.01 & \hphantom{$-$}0.4510 & \hphantom{$-$}0.4527 & 7.27 & 7.23 & 0.5  & 10.79 & 10.65 & 1.3 \\
			& 0.5 & 0.48 & \hphantom{$-$}0.4510 & \hphantom{$-$}0.4953 & 7.27 & 6.98 & 4.1  & 10.79 & 9.70  & 10.1 \\
			& 1   & 0.89 & \hphantom{$-$}0.4510 & \hphantom{$-$}0.5402 & 7.27 & 6.78 & 6.8  & 10.79 & 8.98  & 16.8 \\
			& 2   & 2.27 & \hphantom{$-$}0.4510 & \hphantom{$-$}0.6746 & 7.27 & 6.14 & 15.6 & 10.79 & 6.62  & 38.7 \\
			& 5   & 5.36 & \hphantom{$-$}0.4510 & \hphantom{$-$}0.9220 & 7.27 & 4.95 & 32.0 & 10.79 & 2.22  & 79.4 \\
			\midrule
			
			\multirow{5}{*}{RF}
			& 0   & 0.00 & \hphantom{$-$}0.5765 & \hphantom{$-$}0.5765 & 5.14 & 5.14 & 0.0  & 6.94 & 6.94 & 0.0 \\
			& 0.5 & 0.41 & \hphantom{$-$}0.5765 & \hphantom{$-$}0.5975 & 5.14 & 4.98 & 3.1  & 6.94 & 6.35 & 8.5 \\
			& 1   & 0.84 & \hphantom{$-$}0.5765 & \hphantom{$-$}0.6184 & 5.14 & 4.86 & 5.5  & 6.94 & 5.89 & 15.0 \\
			& 2   & 2.27 & \hphantom{$-$}0.5765 & \hphantom{$-$}0.6843 & 5.14 & 4.52 & 12.2 & 6.94 & 4.62 & 33.3 \\
			& 5   & 5.65 & \hphantom{$-$}0.5765 & \hphantom{$-$}0.8330 & 5.14 & 3.78 & 26.6 & 6.94 & 1.90 & 72.5 \\
			\midrule
			
			\multirow{5}{*}{SVM}
			& 0   & 0.00 & $-$0.0987 & $-$0.0987                  & 7.59 & 7.59 & 0.0  & 13.42 & 13.42 & 0.0 \\
			& 0.5 & 0.85 & $-$0.0987 & \hphantom{$-$}0.0665      & 7.59 & 6.75 & 11.0 & 13.42 & 10.34 & 23.0 \\
			& 1   & 1.77 & $-$0.0987 & \hphantom{$-$}0.2643      & 7.59 & 6.18 & 18.6 & 13.42 & 8.20  & 38.9 \\
			& 2   & 3.22 & $-$0.0987 & \hphantom{$-$}0.5729      & 7.59 & 5.68 & 25.1 & 13.42 & 6.39  & 52.4 \\
			& 5   & 6.27 & $-$0.0987 & \hphantom{$-$}1.2488      & 7.59 & 4.72 & 37.8 & 13.42 & 2.86  & 78.7 \\
			\midrule
			
			\multirow{5}{*}{XGBoost}
			& 0   & 0.00 & \hphantom{$-$}0.5204 & \hphantom{$-$}0.5204 & 6.73 & 6.73 & 0.0  & 10.29 & 10.29 & 0.0 \\
			& 0.5 & 0.48 & \hphantom{$-$}0.5204 & \hphantom{$-$}0.5914 & 6.73 & 6.47 & 3.8  & 10.29 & 9.34  & 9.3 \\
			& 1   & 1.37 & \hphantom{$-$}0.5204 & \hphantom{$-$}0.7032 & 6.73 & 5.91 & 12.2 & 10.29 & 7.25  & 29.5 \\
			& 2   & 2.55 & \hphantom{$-$}0.5204 & \hphantom{$-$}0.8296 & 6.73 & 5.35 & 20.5 & 10.29 & 5.21  & 49.3 \\
			& 5   & 5.91 & \hphantom{$-$}0.5204 & \hphantom{$-$}0.9813 & 6.73 & 4.47 & 33.6 & 10.29 & 1.95  & 81.1 \\
			\bottomrule
		\end{tabular}
		\begin{tablenotes}[flushleft]
			\footnotesize
			\item \textit{RR} denotes the relative FPR reduction with respect to the original decision rule, calculated from unrounded FPR values.
		\end{tablenotes}
	\end{threeparttable}
\end{table}

\subsubsection{FPR within and outside the CP Context}

The temporal CP context contains 2,206 of the 8,128 TWs in the benign variation test set, corresponding to 27.1\% of the evaluated timeline. Table~\ref{tab:fpr_cp_context} compares the FPR within and outside this context. Although the CP context covers only part of the captures, all four models exhibit substantially higher FPRs within these intervals. Outside the context, the FPR ranges from 4.47\% to 5.96\%, whereas the corresponding range within the context is 6.94\% to 13.42\%.

The relative increase ranges from 55.0\% for RF to 147.5\% for SVM. LogReg and XGBoost also show pronounced increases of 81.0\% and 90.4\%, respectively. The magnitude differs across models, but the direction of the change remains consistent. This result indicates that false positives under benign connectivity variations are not distributed uniformly over time but are concentrated in intervals temporally associated with CP activity. The observed concentration motivates applying the adapted threshold only within the CP context while retaining the original decision rule outside it.

\subsection{CP-Aware Adaptation}
\label{subsec:adaptation_results}

Table~\ref{tab:cp_adaptation_tradeoff} presents the test results for the operating points selected with $\delta_{\mathrm{val}}\in\{0,0.5,1,2,5\}$~pp. For each operating point, the selected CP-specific threshold is applied only within the CP context, while the original threshold remains active outside it. Global and CP-context FPRs are measured on the benign variation test set, and $\Delta R_{m,\mathrm{camp}}^{\mathrm{test}}$ denotes the campaign recall reduction on the mixed variation--attack test set. The RR columns report relative FPR reductions with respect to the original decision rule. SVM thresholds are signed decision-function values and may therefore be negative.

Under the original decision rule, campaign recall on the mixed variation--attack test set is 74.55\% for LogReg, 76.38\% for RF, 72.24\% for SVM, and 81.39\% for XGBoost. These values serve as the references for calculating $\Delta R_{m,\mathrm{camp}}^{\mathrm{test}}$. Since the validation tolerance is applied only during adaptation validation, the campaign recall reduction measured on the test set may exceed the corresponding tolerance. The test results are not used to reselect any threshold.

The results show the same general pattern for all four models. Higher values of $\delta_{\mathrm{val}}$ allow higher CP-specific thresholds, which reduce both global FPR and FPR within the CP context but also increase campaign recall reduction. The FPR reduction is greater within the CP context because the adapted threshold is applied only in these intervals. The CP context covers 27.1\% of the benign variation test set, while the original threshold remains active for the other 72.9\% of TWs. The unchanged decisions outside the context therefore limit the reduction in global FPR.

At $\delta_{\mathrm{val}}=0$~pp, RF, SVM, and XGBoost retain their original thresholds. LogReg uses a slightly higher CP-specific threshold of 0.4527 instead of 0.4510. This operating point reduces global FPR by 0.5\% and FPR within the CP context by 1.3\%, while campaign recall decreases by only 0.01~pp on the mixed variation--attack test set.

At $\delta_{\mathrm{val}}=0.5$~pp, global FPR is reduced by 3.1--11.0\%, and FPR within the CP context is reduced by 8.5--23.0\%. The corresponding campaign recall reduction ranges from 0.41 to 0.85~pp. The reduction observed for SVM on the test set exceeds the validation tolerance. At $\delta_{\mathrm{val}}=1$~pp, global FPR reduction reaches 5.5--18.6\%, while FPR reduction within the CP context reaches 15.0--38.9\%. Campaign recall reduction ranges from 0.84 to 1.77~pp. The test-set reductions for SVM and XGBoost exceed the validation tolerance. These results confirm that the validation constraint does not provide a guaranteed bound on the test set. The operating points selected at 2~pp follow the same trend, as reported in Table~\ref{tab:cp_adaptation_tradeoff}.

The largest FPR reductions are obtained at $\delta_{\mathrm{val}}=5$~pp. Global FPR is reduced by 26.6--37.8\%, and FPR within the CP context is reduced by 72.5--81.1\%. These FPR reductions are accompanied by campaign recall reductions of 5.36--6.27~pp. The operating points therefore provide different choices for deployment rather than one universally preferred threshold. Lower tolerances retain more campaign recall, whereas higher tolerances provide larger reductions in false positives temporally associated with benign connectivity variations.

\section{Conclusion and Future Work}
\label{sec:conclusion}

This paper investigated the effect of benign private~5G connectivity variations on frozen IDS models for encrypted OPC~UA traffic. All four models exhibited higher FPRs under connectivity variations than under stable benign conditions. False positives were also more frequent within the temporal CP context, indicating that the increase was concentrated in intervals associated with UE connectivity activity rather than distributed uniformly throughout the captures.

A CP-aware decision adaptation was introduced to address this effect while preserving the underlying traffic-based IDS. Evaluation on the benign variation and mixed variation--attack test captures showed that the method reduces both global FPR and FPR within the CP context. The operating points provide an explicit trade-off between FPR reduction and retained campaign recall. Conservative tolerances preserve most campaign recall while providing moderate FPR reduction, whereas more permissive tolerances achieve substantially larger FPR reductions at the cost of larger campaign recall reductions. The results also show that a recall constraint imposed during validation does not guarantee the same bound on the test data.

The evaluation is limited to four controlled connectivity variations, four OPC~UA attack families, and one industrial private~5G testbed. Future work will consider additional connectivity procedures, network configurations, and industrial applications to assess transferability across operating environments. Further work will investigate attacks and connectivity variations affecting the same UE, automated selection of CP context parameters, and procedure- and intensity-aware adaptations that exploit the individual components of the CP context vector while maintaining explicit constraints on campaign recall.

\section*{Acknowledgment}

This research is funded by dtec.bw -- Digitalization and Technology Research Center of the Bundeswehr. dtec.bw is funded by the European Union -- NextGenerationEU (project ``Digital Sensor-2-Cloud Campus Platform'' (DS2CCP), \url{https://dtecbw.de/home/forschung/hsu/projekt-ds2ccp}).

The authors would like to thank J. Jockram and F. Mueller for their valuable support. We also gratefully acknowledge ipoque GmbH, Deutsche Telekom and Ericsson for their continuous support and cooperation.

	\bibliographystyle{unsrtnat}
	\bibliography{references}
	
\end{document}